\documentstyle[11pt]{article}

\title{\bf On the odderon  intercept in perturbative QCD }
\author{N. Armesto and M. A. Braun$^*$\\
{\it  Departamento de F\'{\i}sica de Part\'{\i}culas,}\\
{\it Universidade de Santiago de Compostela,}\\ 
{\it 15706--Santiago de Compostela, Spain}}
\date{}
\def\beq{\begin{equation}}
\def\eeq{\end{equation}}
\def\noi{\noindent}
\begin{document}
\maketitle
\medskip
\vspace{0.5cm}
\centerline{{\large {\bf Abstract}}}
\begin{quotation}
 Arguments are presented for the odderon intercept being exactly equal to
unity.   A variational method is presented based on a
complete system of one-gluon functions. For the odderon, the highest intercept
calculated by this method is $1-(3\alpha_{s}/\pi)\,0.45.$
\end{quotation}
\vspace{8.0 cm}

\noi{\large March 1996}

\noi{\large \bf hep-ph/9603218}

\noi{\Large\bf US-FT/7-96}

\vspace{0.5cm}
\noi$^*$ Visiting Professor IBERDROLA. On leave of absence from the Department 
of High Energy
Physics, University of St. Petersburg, 198904 St. Petersburg, Russia.

 \newpage

\section{Introduction}

\ \ \ \ Much attention has recently been devoted to the perturbative "hard", or
BFKL, pomeron \mbox{[1]}, especially in relation to the study of the small $x$
behaviour 
 of the deep inelastic scattering structure functions (see a recent review 
in [2]). In application to soft phenomena, the value of the pomeron
intercept is of principal importance. For the BFKL pomeron it is
considerably above unity: 
\[ \alpha_{BFKL}(0)=1-(3\alpha_{s}/\pi)E_{0},\]
where the "energy" $E_{0}$ is equal to $-4\ln 2$, and $\alpha_{s}$ is the
(fixed) QCD coupling constant [1]. As a result, to obtain a unitary amplitude
one has to take into account more than two, in fact, any number of interacting
reggeized gluons. This problem simplifies in the large $N_{c}$ limit, when
it reduces to summing all multipomeron exchanges [3].

For the negative signatured amplitude the lowest order contribution comes
from the exchange of an odderon, a state formed from three reggeized gluons
in a symmetric colour configuration [4]. Its relative importance is
controlled by the odderon intercept. Should it also lie above unity, 
unitarization would require summing any number of exchanged odderons as well.

It has not been possible to obtain a complete solution of the odderon
equation ("the BKP equation") up to now. Certain encouraging ideas have
been proposed, however, in [5-7], based on the conformal symmetry of the
equation and the Yang-Baxter technique. In [8] the conformal symmetry was
used to reduce the problem to an one-dimensional equation. Variational
calculations on the basis of this approach, with a relatively simple trial
function containng two free parameters, gave an intercept
above unity, although lower than for the pomeron [9]:
\[\alpha_{odd}(0)=1-(3\alpha_{s}/\pi)E_{odd},\ E_{odd}<-0.37.\]

A different scheme of variational calculations of the odderon intercept was
adopted in our paper [10]. We studied the odderon energy matrix in a non
conformally invariant basis of functions, whose number was taken rather large
(up to 3368), which corresponds to a correspondingly large number of
variational parameters. Our best result for the energy  however turned
out positive \[E_{odd}<0.45\]
corresponding to an intercept below unity. In view of the variational
character of the calculations this result evidently does not contradict [9]
but seems to be much weaker. For that reason we did not give much importance
to our result  at the time when it was obtained, so that it remained
unpublished.

However, further study of the odderon problem has given us some motivation to
believe that this result may be closer to reality than the one obtained in
[9]. The point is that the equation for the odderon at rest admits a simple
solution corresponding to the energy exactly equal to zero and the intercept
$j=1$. This solution is based on the so-called bootstrap relation in the BFKL
theory [11] and we call it the bootstrap solution. One can view this
solution as a true bound state (normalizable). It is nodeless and possesses
a maximal possible symmetry. Therefore, from the experience gained in
quantum-mechanical problems, we can expect it to correspond to the ground
state of the odderon, that is, to the lowest energy possible.

In Sec. 2 we discuss this point in more detail giving some mathematical
arguments in favour of this result.
In view of this, our calculations aquire
a better status, which gives us a reason to  present them in this
paper, in Secs. 3-5. Some  conclusions are drawn in Sec. 6.
All mathematical details are relegated to  Appendices.

\section{The bootstrap solution as the ground state of the odderon}

\ \ \ \ As shown in [4] the odderon wave function in the transversal
momentum space $\psi (q_{1},q_{2},q_{3})$ satisfies a Schroedinger-like
equation
\beq
H\psi=E\psi.
\eeq
The hamiltonian $H$ is a sum of kinetic terms and pair interactions between
the three gluons:
\beq
H=T_{1}+T_{2}+T_{3}+U_{12}+U_{23}+U_{31}.
\eeq
Each kinetic term is given by the gluonic Regge trajectory with a minus sign
in units $3\alpha_{s}/\pi$: \beq
T_{1}=T(q_{1})=-\omega(q_{1})=\frac{\eta(q_{1})}{4\pi}
\int \frac{ d^{2}q'_{1}}{\eta(q'_{1})\eta(q_{1}-q'_{1})},
\eeq
where, with an infrared regularization provided by the gluon mass $m$,
\beq
\eta(q)=q^{2}+m^{2}.
\eeq 
 The interaction terms $U$ are
integral operators in the momentum space of the three gluons with an
integration measure
\beq
d\mu=\frac{d^{2}q_{1}d^{2}q_{2}d^{2}q_{3}\delta^{2} (q_{1}+q_{2}+q_{3}-q)}
{\eta(q_{1})\eta(q_{2})\eta(q_{3})}
\eeq
where $q$ is the (fixed) total  transverse momentum of the odderon. In this
section we shall consider only the case $q=0$ so that $q_{1}+q_{2}+q_{3}=
0$ will be  assumed. Explicitly the kernel, say, of $U_{12}$ is given
by
\beq
U_{12}(q_{1},q_{2},q_{3}|q'_{1},q'_{2},q'_{3})=
\eta(q_{3})\delta^{2}(q_{3}-q'_{3})V_{12}(q_{1},q_{2}|q'_{1},q'_{2}),
\eeq
where $V_{12}$ is the BFKL interaction for two gluons in a vector colour
state: \beq
V_{12}(q_{1},q_{2}|q'_{1},q'_{2})=-\frac{1}{4\pi}
\left [\frac{\eta_(q_{1})\eta_(q'_{2})+\eta_(q'_{1})\eta_(q_{2})}
{\eta(q_{1}-q'_{1})}-\eta(q_{1}+q_{2})\right ].
\eeq
(Note that this interaction is twice smaller than the one for the vacuum
chanel which enters the standard pomeron equation).

It is well-known that like the pomeron equation, the BKP equation (1) 
is infrared stable, that is, it remains meaningful if one puts $m=0$ in (4)
[4]. However, for our purpose it will be convenient to proceed in a
different manner. Rescaling $q\rightarrow q/m$ we eliminate $m$ from Eq. (1)
altogether, the resulting $\eta(q)$ having the form
\beq
\eta(q)=q^{2}+1.
\eeq
Thus one observes that Eq. (1) (homogeneous!) is in fact independent of $m$
and so are the energy eigenvalues $E$, which is of no wonder, since they are
dimensionless. In the following we shall discuss Eq. (1) in the infrared
regularized form provided by $\eta$ given by (8). It clearly shows that the
singularities of the equation come only from the ultraviolet region when
$q_{1,2,3}\rightarrow\infty$.

The crucial point for our discussion is the existence of the bootstrap
solution which is a constant:
\beq
\psi_{B}(q_{1},q_{2},q_{3})=\psi_{0}.
\eeq
As mentioned, the bootstrap identity is essential for this:
\beq
\int\frac{d^{2}q'_{1}}{\eta(q'_{1})\eta(q'_{2})}V_{12}
(q_{1},q_{2}|q'_{1},q'_{2})=\omega(q_{1})+\omega(q_{2})-\omega(q_{1}+q_{2}),
\eeq
which is a consequence of the structure of $V$ and $\omega$ in terms of $\eta$
and valid for an arbitrary $\eta(q)$.
Applying this identity we find
\[ H\psi_{B}=\psi_{0}(-\omega(q_{1})-\omega(q_{2})-\omega(q_{3})\]\[
+2\omega(q_{1})+2\omega(q_{2})+2\omega(q_{3})-
\omega(q_{1}+q_{2})-\omega(q_{2}+q_{3})-\omega(q_{3}+q_{1}))=0.\]
Here the first three terms come from the kinetic part and the rest from the
interaction according to (10). This expression vanishes because at $q=0$ we have
$\omega (q_{1}+q_{2})=\omega(q_{3})$ etc. Evidently the solution (9) is
normalizable with the measure given by (5) and $\eta$ taken according to (8):
\beq
\Vert \psi_{B}\Vert^{2}=|\psi_{0}|^{2}\int d\mu<\infty.
\eeq
Therefore it represents a true bound state of the three gluons with a zero
energy. It evidently possesses the maximal symmetry possible. As mentioned,
our experience with quantum-mechanical problems then suggests that it is the
ground state of the system.

To somewhat strengthen this proposition, we take a bit more mathematical
point of view. The integral equation (1) is singular for two reasons. One is
evident and comes from the pairwise nature of the interaction. It is common to
all quantum mechanical three-body problems. As is well-known, it can be cured
by going over to the equivalent  Faddeev matrix equations for parts of the
wave function.
However  there is another source of the singularity related to a
bad ultraviolet behaviour of the kinetic terms: they grow very slow with
$q\rightarrow\infty$ (only logarithmically) making the equation badly
singular. This singularity persists also in the corresponding two-body
equation for the BFKL pomeron and is responsible for its spectrum to be
different from the  "free" equation with $V=0$. To cure this singularity we
introduce an ultraviolet cutoff into $\eta$ substituting (8) by
\beq
\eta_{\nu}(q)=\eta(q)\exp (\nu q^{2}),\ \ \ \nu >0
\eeq
With $\nu>0$ the kinetic terms grow exponentially with the momentum. One
easily finds that
\beq
\omega_{\nu}(q)_{q\rightarrow\infty}\simeq -
\frac{2}{\nu q^{2}}\exp (\frac{1}{2}\nu q^{2}).
\eeq
Now one can apply the standard methods to study the spectrum of the odderon
regularized in the ultraviolet.

Let us illustrate how it works with a much simpler two-body problem of the
pomeron at rest. After changing its wave function $\psi(q_{1},q_{2})$ with
$q_{1}+q_{2}=0$ according to
\[\psi(q_{1},q_{2})=\eta_{\nu} (q_{1})\phi(q_{1})\]
the pomeron equation reduces to
\beq
(H_{0}(q)-E)\phi(q)=-\int d^{2}q'U(q,q')\phi (q'),
\eeq with
$H_{0}(q)=-2\omega_{\nu}(q)$ and the kernel $U$ given by
\beq
U(q,q')=-\frac{1}{2\pi}
(\frac{2}{\eta_{\nu}(q-q')}-\frac{\eta_{\nu}(0)}
{\eta_{\nu}(q)\eta_{\nu}(q')})
\eeq
From (14) we go over to an equivalent eigenvalue equation:
\beq
K(E)\chi_{\lambda}=\lambda (E)\chi_{\lambda},
\eeq
where
\beq
K(E)=(H_{0}-E)^{-1/2}U(H_{0}-E)^{-1/2}.
\eeq
Evidently the energy eigenvalues $E$ for Eq. (14) are determined from the ones
for (16) by the equation
\beq
\lambda(E)=-1
\eeq
and the corresponding eigenfunctions are related by
\beq
\phi=(H_{0}-E)^{-1/2}\chi_{-1}.
\eeq
Eq. (16) has usually much better properties as compared to the initial
Schroedinger equation (14). The denominators in (17) normally make the kernel
$K$ to be of the Fredholm type, provided that both $H_{0}$ and $U$ are not too
badly behaved and that $E$ does not lie inside the spectrum of $H_{0}$. Then
Eq. (16) possesses only a discrete spectrum of eigenvalues, which may be
ordered according to their absolute values as
\[ |\lambda_{1}|>|\lambda_{2}|>....\]
With a finite norm of $K$, the number of $\lambda$'s with an absolute value
more than unity is finite.  According to (18) this means that only a finite
number of discrete eigenvalues $E$ may exist below the continuum spectrum,
the latter coinciding with that of $H_{0}$. 

In the pomeron case we find by a direct calculation that with $\nu>0$ 
the norm of the kernel $K(E)$ is finite for all $E$ below the minimal
value of $H_{0}$:
\beq
\Vert K_{\nu}(E)\Vert^{2}=\int d^{2}qd^{2}q'|K_{\nu}(E;q,q')|^{2}<\infty.
\eeq
One can even calculate the norm at $E=0$ and $\nu\rightarrow\infty$:
\beq
\Vert K_{\nu}(0)\Vert^{2}_{\nu\rightarrow\infty}=64/9.
\eeq
Note that this limit corresponds to $q^{2}\sim 1/\nu$ and consequently to an
$\eta(q)$ in the form of a pure exponential
\beq
\eta_{\nu}(q)_{\nu\rightarrow\infty}\simeq\exp(\nu q^{2}).
\eeq

What do these results tell us about the physically interesting 
pomeron spectrum at $\nu\rightarrow 0$? Very little indeed. One observes that
with a large $\nu$ there can be at most 7 eigenvalues $\lambda_{i}$ with
an absolute value greater than unity at $E=0$. This means that there are at
most 7 discrete negative energy levels  at large $\nu$. As $\nu$ diminishes
the norm of $K$ becomes larger and  the number of discrete negative
levels $E$ also increases. When $\nu\rightarrow 0$ the norm blows up to
infinity and so does the number of negative energy levels. The latter finally
form a cut which completely fills the gap between the BFKL energy level
$E_{BFKL}=-(12\alpha_{s}\pi)\ln 2$ and the kinetic energy threshold
$E_{0}=3\alpha_{s}/2\pi$. Except for this behaviour with the change of
$\nu$, no new information can be extracted from this approach in the pomeron
case.

However, applied to the odderon case this argument leads to  certain
important consequences. One notices that with $\eta$ changed to $\eta_{\nu}$
the bootstrap identity (10) remains valid and thus the bootstrap solution
persists with the energy eigenvalue $E=0$  at any $\nu$. This means that
for the odderon equation  similar to (16) a curve $\lambda_{B}(E)$ always
exists which passes through minus unity at $E=0$: $\lambda_{B}(0)=-1$.
Since different $\lambda$'s cannot cross with changing $\nu$ and since
$|\lambda |$'s are expected to fall monotonously with $|E|$ for $E$ below the
spectrum of $H_{0}$, we find that the curve $\lambda_{B}(E)$ divides all
$\lambda$'s into two separate sets: those which lie above it and those which
lie below it (Fig. 1), and that this division is conserved as $\nu$ changes.
Since we expect the norm of the operator $K$ in the properly formulated
(Faddeev) equation to be finite at $E=0$, the number of $\lambda$'s below
$\lambda_{B}$ has also to be finite and it is conserved with the changing
$\nu$. Then, as a first result, we find that the solutions of the physical
odderon equation with $\nu=0$ with negative  energies form a discrete finite
set. A second result is that the number of such solutions can be studied at
any chosen value of $\nu$, in particular at large $\nu$, since this number is
adiabatically conserved with $\nu$. The odderon theory with a large $\nu$
and the exponential $\eta$ of the form (22) is much simpler than for the
physical odderon with $\nu=0$. Direct variational estimates reveal that in
this case there are no solutions with nonpositive energies $E$ at large $\nu$
except the bootstrap one (see Appendix 1.). This brings us to the conclusion
that the bootstrap solution is indeed the one with the lowest energy.

Of course, this argument is not absolutely rigorous. To make it such, one has
to find the norm of the corresponding Faddeev kernel (a $3\times 3$ 
{\mbox skew-
symmetric} matrix of two-body scattering matrices). Before that one has to
study these pair scattering matrices and show that their properties are no
worse than those of the interactions $U_{ik}$ given by (6) and (7). This seems 
realizable although rather difficult. Also, even knowing that the norm of $K$
is finite, one cannot exclude in principle some bizarre behaviour of the
eigenvalues $\lambda$ which might invalidate the above logic. For that reason
we do not consider our derivation as a definite proof but rather as a strong
argument in favour of the bootstrap solution to represent the odderon ground
state.

 \section{Variational calculations of the  odderon ground
state energy }

\ \ \ \ To perform variational estimates of the odderon ground state energy
it is more convenient to set $m=0$ from the start, to simplify the explicit
form of the Hamiltonian. To make it non-singular we pass to a new wave
function by a substitution  
\beq
\psi\rightarrow\prod_{i=1}^{3}q_{i}^{2}\psi.
\eeq
We shall not fix the total momentum of the odderon. Then for the new 
$\psi$ and
with $m=0$ the metric (5) changes to
\beq
d\mu=\prod_{i=1}^{3}q_{i}^{2}d^{2}q_{i}.
\eeq
The odderon equation (1) becomes
\beq
H\psi=E\prod_{i=1}^{3}q_{i}^{2}\psi,
\eeq
 The Hamiltonian can be written as [ 8 ]
\beq
H=(1/2)(H_{12}+H_{23}+H_{31})
\eeq
where $H_{ik}$ is the BFKL Hamiltonian for gluons $i$ and $k$ in units
$3\alpha_{s}/2\pi$, which in the  limit
$m=0$ acts on the wave function as
\beq
H_{ik}\psi=\prod_{j=1}^{3}q_{j}^{2}(\ln q_{i}^{2}q_{k}^{2}+4{\bf C})\psi
+\prod_{j=1,j\neq i,k}^{3}q_{j}^{2}\left [q_{i}^{2}\ln
(r_{ik}^{2}/4)\,q_{k}^{2} +(i\leftrightarrow
k)\right ]\psi+2(q_{i}+q_{k})^{2}\psi(r_{ik}=0), \eeq
where $r_{ik}=r_{i}-r_{k}$ is the (transversal) distance between the gluons
and $\bf C$ is the Euler constant.

The solution of (25) may  be found by a variational approach,
searching
 the minimum value of the functional
\beq
\Phi=\int\prod d^{2}q_{i}\psi^{\ast}H\psi,
\eeq
with the normalization condition
\beq
\int\prod d^{2}q_{i}\psi^{\ast}\prod_{i=1}^{3}q_{i}^{2}\psi=1.
\eeq
Using the symmetry of the wave functions in the three gluons, one can study a
simpler functional
\beq
{\cal E}=(1/2)\int\prod_{i=1}^{3}d^{2}q_{i}\psi^{\ast}H_{12}\psi.
\eeq
The odderon energy
is determined by the minimal value  $\epsilon_{3}$ of $\cal E$  according
to \beq
E_{odd}=(3/2)\epsilon_{3}.
\eeq
Note that the minimal value $\epsilon_{2}$ of the same functional on functions
$\psi(q_{1},q_{2})$ for two gluons determine the pomeron energy:
$E_{0}=\epsilon_{2}$.

We expand the  wave
function in a sum of products of individual gluon functions: 
\beq
\psi(r_{1},r_{2}, 
r_{3})=\sum_{\alpha_{1},\alpha_{2},\alpha_{3}}
c_{\alpha_{1},\alpha_{2},\alpha_{3}}\prod
_{i=1}^{3}
\psi_{\alpha_{i}}(r_{i}),
\eeq
where the one-gluon functions $\psi_{\alpha}(r_{i})$ form a discrete
complete set and are orthonormalized according to (29):
\beq
\int d^{2}r\psi_{\alpha}^{\ast}q^{2}\psi_{\alpha'}=\delta_{\alpha,\alpha'}.
\eeq
The coefficients $c_{\alpha_{1},\alpha_{2},\alpha_{3}}$ have to be symmetric
in
all $\alpha$'s by the requirement of the Bose symmetry and normalized
according to \beq
\sum_{\alpha_{1},\alpha_{2},\alpha_{3}}
|c_{\alpha_{1},\alpha_{2},\alpha_{3}}|^{2}=1.
\eeq
The two-gluon
Hamiltonian
$H_{12}$ acts nontrivially only on the wave functions for the gluons  1 and 2.
 So the energy functional becomes
\beq
{\cal
E}=\sum_{\alpha_{1},\alpha_{2},\alpha'_{1},\alpha'_{2},\alpha_{3}}
c^{\ast}_{\alpha_{1},\alpha_{2},\alpha_{3}}
c_{\alpha'_{1},\alpha'_{2},\alpha_{3}}{\cal
E}_{\alpha_{1},\alpha_{2},\alpha'_{1},\alpha'_{2}}, \eeq
where the matrix ${\cal E}_{\alpha_{1},\alpha_{2},\alpha'_{1},\alpha'_{2}}$
is the two-gluon energy  in the basis formed by functions $\psi_{\alpha}$.
With this matrix known, the problem of  minimization of the functional
$\cal E$ reduces to finding the minimal value of a cuadratic form, that is,
 the minimal eigenvalue of the matrix ${\cal
E}_{\alpha_{1},\alpha_{2},\alpha'_{1},\alpha'_{2}}$ considered as a matrix in 
independent initial and final $3$-gluon states. The latter means that this
matrix should be multiplied by unity matrix for  the third gluon
and then symmetrized in all initial and final gluons. The procedure is quite
straightforward, once the basic functions $\psi_{\alpha}$ are chosen. It
however involves a numerical evaluation of the energy matrix elements and a
diagonalization of the matrix, whose dimension is rapidly growing with the
number of  the basic functions taken into account.

\section{Two-gluon energy matrix for given angular momenta}

\ \ \ \ Introducing the individual gluon angular momentum $l$ we take
$\phi$ trivially:
\beq
\psi_{\alpha}({\bf r})=\psi_{k,l}(r)\exp il\phi,
\eeq
where $k=0,1,2,...$ enumerates the radial functions.
 In the following, instead
of $r$, we shall use the variable $z=\ln r^{2}$ in most cases. In terms of
$z$ and $\phi$
\beq
q^{2}\psi_{k,l}({\bf r})
=-(4/r^{2})(\partial^{2}_{z}-(1/4)l^{2})\psi_{k,l}({\bf r}).
\eeq
Wave functions with different values of the angular momentum are
automatically orthogonal. For coinciding $l$  the normalization
condition for the radial functions  reduces to the standard form for functions
\beq
\xi_{k,l}(z)=(\partial+|l|)\psi_{k,l}(z),
\eeq
which should satisfy
\beq
\int dz \xi^{\ast}_{k,l}(z)\xi_{k',l}(z)=(1/4\pi )
\delta_{kk'}.
\eeq
 We assume that the radial functions are chosen to be real.

With the angular dependence of the wave function explicitly given by (36), one
can do the azymuthal integrals in the potential energy in a straightforward
manner. Let $\alpha_{i}=\{k_{i},l_{i}\}$ and take the transition between two
gluon states
$\alpha_{1},\alpha_{2}\rightarrow\alpha_{3},\alpha_{4}$.
 Evidently the total
angular momentum is conserved so that the energy matrix elements are zero
unless
$ l_{1}+l_{2}=l_{3}+l_{4}$.
 According to (27) the potential energy consists of
two parts, the first
 part $U$ given by an essentially Coulomb interaction and
the second one $Q$ given by a contact interaction, proportional to their
total momentum squared. Let us begin with the Coulomb part $U$. Its two terms
evidently give the same contribution due to the symmetry under the
interchange of gluons 1 and 2. So we can take only one of them and drop
the factor $1/2$. Denote
\beq
\eta_{k,l}(z)=(\partial^{2}-(1/4)l^{2})\psi_{k,l}(z).
\eeq
Then after doing the azymuthal integration we obtain 
\beq
U_{\alpha_{1},\alpha_{2};\alpha_{3},\alpha_{4}}=16\pi^{2}
\int dz_{1}dz_{2}\eta_{\alpha_{1}}(z_{1})\psi_{\alpha_{3}}(z_{1})
\psi_{\alpha_{2}}(z_{2})\eta_{\alpha_{4}}(z_{2})U_{l}(z_{1},z_{2}),
\eeq
where
$l=|l_{1}-l_{3}|=|l_{2}-l_{4}|$ 
is the angular momentum transfer and the function $U_{l}(z_{1},z_{2})$
is given by
\beq
U_{l}=-(1/l)\exp (-(l/2)|z_{1}-z_{2}|), \ \ l\neq 0,
\eeq
and
\beq
U_{0}=\max \{z_{1},z_{2}\}.
\eeq

The contact part $Q$ involves gluonic wave functions taken at the same point.
 After performing the azymuthal
integration and integrating once by parts in the variable $z$ we obtain
\beq
Q_{\alpha_{1},\alpha_{2};\alpha_{3},\alpha_{4}}=8\pi^{2}
\int dz \left [(\partial
+(1/2)|l_{1}+l_{2}|)\psi_{\alpha_{1}}\psi_{\alpha_{2}}\right ]
 \left [(\partial
+(1/2)|l_{3}+l_{4}|)\psi_{\alpha_{3}}\psi_{\alpha_{4}}\right ]. \eeq
One can somewhat simplify this expression by noting that
\beq
(\partial +(1/2)|l_{1}+l_{2}|)\psi_{\alpha_{1}}\psi_{\alpha_{2}}=
\xi_{\alpha_{1}}\psi_{\alpha_{2}}+\psi_{\alpha_{1}}\xi_{\alpha_{2}}
+\Delta_{12}\psi_{\alpha_{1}}\psi_{\alpha_{2}},
\eeq
where $2\Delta_{12}=|l_{1}+l_{2}|-|l_{1}|-|l_{2}|$ and similarly for the second
factor in (44). Then finally
\beq
Q_{\alpha_{1},\alpha_{2};\alpha_{3},\alpha_{4}}=8\pi^{2}
\int dz (\xi_{\alpha_{1}}\psi_{\alpha_{2}}+\psi_{\alpha_{1}}\xi_{\alpha_{2}}
+\Delta_{12}\psi_{\alpha_{1}}\psi_{\alpha_{2}})
(\xi_{\alpha_{3}}\psi_{\alpha_{4}}+\psi_{\alpha_{3}}\xi_{\alpha_{4}}
+\Delta_{34}\psi_{\alpha_{3}}\psi_{\alpha_{4}}).
\eeq

The kinetic energy is easier to calculate in the momentum space. So we
transform the basic functions to the momentum space according to
\beq
\psi_{\alpha}({\bf q})=\int (d^{2}r/2\pi)\psi_{\alpha}({\bf r})\exp
  (-i{\bf qr})=(-i)^{l}\exp il\phi
\,\int rdr\psi_{k,l}(z)J_{l}(qr).
\eeq
where $J_{l}$ is the Bessel function. To do the integral over $r$ it is
convenient to introduce a Fourier transform of the function $\psi$ with
respect to the variable $z$:
\beq
\psi_{k,l}(z)=\int (d\nu/\sqrt{2\pi})\phi_{k,l}(\nu)\exp i\nu z.
\eeq
Putting this representation in (47) and doing the $r$-integration we obtain
\beq
\psi_{k,l}({\bf q})=(2/q^{2})\exp il\phi
\int (d\nu/\sqrt{2\pi})f_{k,l}(\nu)q^{-2i\nu},
\eeq
with
\beq
f_{k,l}(\nu)=(-i)^{|l|}2^{2i\nu}(|l|/2+i\nu)\phi_{k,l}(\nu)
\Gamma (|l|/2+i\nu)/\Gamma(|l|/2-i\nu).
\eeq

With the gluon wave functions in the momentum space given by (49), both
radial and azymuthal integration in $\bf q$ are easily done. The final matrix
element of the kinetic energy $T$ results as
\beq
T_{\alpha_{1},\alpha_{3}}=-4\pi i\int d\nu f^{\ast}_{k_{1},l_{1}}(\nu)
(\partial/\partial\nu )f_{k_{3},l_{1}}(\nu),\ \ l_{1}=l_{3}
\eeq
The differentiation gives
\beq
(\partial/\partial\nu )f_{k_{3},l_{1}}(\nu)=f_{k_{3},l_{1}}(\nu)
\left [ 2i\ln 2+2i{\mbox Re}\, \psi (|l_{1}|/2+i\nu)+
(\partial/\partial\nu )\ln \left [(|l|/2+i\nu)\phi_{k,l}(\nu)\right ]
\right]. \eeq
Correspondingly the kinetic energy matrix element separates into terms
\beq
T^{(1)}_{\alpha_{1},\alpha_{3}}=8\pi\int d\nu f^{\ast}_{k_{1},l_{1}}(\nu)
f_{k_{3},l_{1}}(\nu)\left [\ln 2+{\mbox Re}\, \psi (|l_{1}|/2+i\nu)\right ]
\eeq
and
\beq
 T^{(2)}_{\alpha_{1},\alpha_{3}}=-4\pi i\int d\nu 
\left [
(|l_{1}|/2+i\nu)\phi_{k_{1},l_{1}}(\nu)\right]^{\ast}(\partial/\partial\nu )
\left [(|l_{1}|/2+i\nu)\phi_{k_{3},l_{1}}(\nu)\right ]. \eeq
The function $(|l|/2+i\nu)\phi_{k,l}(\nu)$ is nothing but
the Fourier transform  of $\xi_{k,l}(z)$ with respect to $z$.
Correspondingly we denote it as
\beq
(|l|/2+i\nu)\phi_{k,l}(\nu)\equiv \xi_{k,l}(\nu).
\eeq
The part $T^{(2)}$ can then be written as
\beq
 T^{(2)}_{\alpha_{1},\alpha_{3}}=-4\pi i\int d\nu 
\xi_{k_{1},l_{1}}(\nu)^{\ast}(\partial/\partial\nu )
\xi_{k_{3},l_{1}}(\nu).
\eeq

The orthonormalization property (39)  transforms into the analogous
property in the $\nu$ space
\beq
\int d\nu \xi^{\ast}_{k,l}(\nu)\xi_{k',l}(\nu)=(1/4\pi)\delta_{kk'}.
\eeq
Noting that
$f^{\ast}_{k,l}(\nu)f_{k',l}(\nu)
=\xi^{\ast}_{k,l}(\nu)\xi_{k',l}(\nu)$ we observe
 that the term $\ln 2$ in (53) will add a constant $2\ln 2$ to
the energy. Separating another constant term $2\psi(1)$ we finally present
the part $T^{(1)}$ in the final form
\beq
T^{(1)}_{\alpha_{1},\alpha_{3}}=2(\ln 2+\psi(1))\delta_{\alpha_{1},\alpha_{3}}
+8\pi\int d\nu\xi^{\ast}_{\alpha_{1}}(\nu)\xi_{\alpha_{3}}(\nu)
\left [{\mbox Re}\, \psi (|l_{1}|/2+i\nu)-\psi(1)\right ].
\eeq
The first, constant, term cancels an identical one in the initial Hamiltonian
(27). 
Using the representation  
\beq
\psi(x)-\psi(1)=\int_{0}^{\infty}dt\left
[\exp(-t)-\exp(-xt)\right ]/(1-\exp(-t)) \eeq
and the othornormalization property of the set $\xi_{\alpha}$ we may cast
 $T^{(1)}$ in the form
\beq 
T^{(1)}_{\alpha_{1},\alpha_{3}}=
2\int_{0}^{\infty}(dt/(\exp t-1))(\delta_{\alpha_{1}\alpha_{3}}-
\exp(t(1-|l_{1}|/2))g_{\alpha_{1}\alpha_{3}}(t)),
\eeq
where
\beq
g_{\alpha_{1}\alpha_{3}}(t)=4\pi\int d\nu
\xi^{\ast}_{\alpha_{1}}(\nu)\xi_{\alpha_{3}}(\nu)\cos \nu t.
\eeq
Note that (59) is not valid for ${\mbox Re}\,x=0$. Therefore  this formula
cannot be applied when the gluon orbital momentum is zero. In this case one
may use
\[ \psi(i\nu)+\psi(-i\nu)=\psi(1+i\nu)+\psi(1-i\nu),\]
which formally corresponds to changing the angular momentum to be equal to 2.

As to the second part of the kinetic energy $T^{(2)}$, it turns out to
 be cancelled by a similar contribution coming from the
monopole part of the Coulomb interaction for the angular momentum transfer
equal to zero (see Appendix 2.).

 Most of the contributions to the energy presented in this
section can hardly be further simplified and were used in the numerical
calculations as they stand. The exception is the monopole part of the Coulomb
interaction corresponding to (41) with $l=0$ (Eq. (43)). This part contains
contributions which cancel the term $T^{(2)}$ in the kinetic energy and
partially the contact interaction contribution for $l=0$. The cancellation
between the monopole Coulomb interaction and the kinetic term $T^{(2)}$ is
responsible for the scale invariance of the energy. Calculation of the
monopole Coulomb part is discussed in  Appendix 2.

\section{The choice of basic functions and the numerical results}

\ \ \ \ A natural orthonormal discrete basis for $z=\ln r^{2}$ varying from
$-\infty$ to $+\infty$ is formed by the harmonic oscillator proper functions.
Thus we choose functions $\xi_{k,l}(z)$ independent of $l$ and given by
\beq
\xi_{k}(z)=c_{k}H_{k}(z)\exp (-z^{2}/2),
\eeq
where $H_{k}$ are the Hermite polinomals and $c_{k}$ are determined by the
normalization condition (39) to be
\beq
c^{2}_{k}=1/(4\pi^{3/2}2^{k}k!).
\eeq
The  Fourier transformation to the $\nu$ space gives
\beq
\xi_{k}(\nu)=(-i)^{k}c_{k}H_{k}(\nu)\exp (-\nu^{2}/2).
\eeq  
In the coordinate space the function $\eta_{k,l}(z)$ is obtained from $\xi$
by differentiation:
\beq
\eta_{k,l}(z)=(\partial-(1/2)|l|)\xi_{k}(z).
\eeq
The function $\psi_{k,l}$ is obtained from $\xi_{k}$ as a solution of the
 differential equation
\beq
\xi_{k}(z)=(\partial+(1/2)|l|)\psi_{k,l}(z),
\eeq
with a boundary condition $\psi_{k,l}(-\infty)=0$. It is given by an integral
\beq
\psi_{k,l}(z)=\int_{-\infty}^{z}dz'\xi_{k}(z')\exp(-|l|(z-z')/2).
\eeq

With this set of functions the potential part of the energy was calculated
numerically. As to the kinetic part, the function $g$ entering (60) can be
found analytically. For transition $k,l\rightarrow k',l$ it is equal to zero if
$k+k'$ is odd. For even $k+k'=2s$ and $k\geq k'$
\beq
g_{kk'}(t)=4\pi^{3/2}(-1)^{d}c_{k}c_{k'}\exp(-t^{2}/4)\sum_{p=0}^{k'}
2^{p}p!C_{k}^{p}C_{k'}^{p}(-t^{2})^{s-p},
\eeq
where $2d=k-k'$.

After the energy matrix ${\cal
E}_{\alpha_{1},\alpha_{2},\alpha'_{1},\alpha'_{2}}$ is calculated
and properly symmetrized in the three gluons,  its lowest
eigenvalue is determined, which gives an upper limit on the exact
 odderon energy according to Eq. (31). To study the minimal energy
only states with the total angular momentum equal to zero have been included.

The selected set of basic one-gluon functions is characterized by the maximal
value of the angular momentum included $l_{max}$ and numbers of radial
functions included for each wave. As calculations show, best results are
obtained when one raises $l_{max}$ and the number of radials in all waves
simultaneously. So we present here the results for the case when the number
of radials $r$ is the same for all angular momenta and is equal to the number
of angular momenta included: $r=l_{max}+1$. Such a set of functions is thus
characterized by a single parameter $r$. With a growth of $r$ the number of
states $N$ rises very rapidly. In our calculations the number $r$ was limited
to 6.

To study convergence of the procedure, it was first applied to the pomeron,
with only two gluons, where the exact energy is known. The  maximal $r=6$
corresponds to 201 basic two-gluon states in this case.  For the odderon  with
$r=6$ the number of basic states rises to 3368.

The results of the calculations of the ground state energies of the BFKL
pomeron $\epsilon_{2}$ and the odderon $\epsilon_{3}$ are presented in the
Table  for different values of $r$.
 One observes that for the pomeron  the obtained energies are still rather far
from the exact value. 
 Thus the convergence of the method is rather slow. The Table
also reveals
that the odderon energy is essentially
larger than the pomeron one for a given $r$. So our results confirm
that, in all probability, the odderon intercept is lower than that of the BFKL
pomeron. Our values for the odderon energy  stay positive, approaching 
zero quite slowly. As discussed in Sec. 2, this seems to favour the minimal
odderon energy being equal to zero.

\section{Conclusions}

We have presented arguments that the odderon intercept is  exactly
equal to unity, starting from the explicitly known bootstrap solution.
This solution corresponds to the odderon at rest. However for lack of scale,
the intercept should be the same for nonzero momentum transfers. Our
variational calculations seem to confirm this result.

If it is correct then experimentally we would expect the difference between
the cross-sections on a particle and its antiparticle to stay constant at
high energies. Also the odderon effects in the cross-sections themselves
should decrease with energy, since the pomeron contribution rises with it.

On the theoretical side, the intercept of the odderon being equal to one
means that exchanges of more than one odderon are of no importance. Then the
unitarization program essentially reduces to summing only pomerons. As
mentioned this can be realized in the high-colour limit. However to do it for
the negative signatured amplitude, one has to know how to couple the odderon
to an external sourse perturbatively. This seems to be an unsolved problem.

\section{Acknoledgements}

\ \ \ \ The authors are grateful to the  Direccion General
de Investigacion Cientifica y Tecnica (DGICYT) of Spain and to the
Xunta de Galicia for financial support.

\section{Appendix 1. Variational estimates of the odderon ground state energy
with an exponential $\eta$}

\ \ \ \ With (22) the form of the kinetic part and the interaction, as well
as that of the measure (5), simplify drastically and reduce to Gaussians.
Then one can calculate the average energy quite easily, provided the trial
function is also taken Gaussian or a sum of Gaussians. We performed
variational estimates of the ground state energy choosing the trial function
in the form
\beq
\psi(q_{1},q_{2},q_{3})=\sum_{i}c_{i}e^{-\beta_{i}(q_{1}^{2}+q_{2}^{2}+
q_{3}^{2})},\ \ q_{1}+q_{2}+q_{3}=0,
\eeq
with variational parameters $c_{i}$ and $\beta_{i}\geq 0$.

The norm of $\psi$ turns out to be
\beq
\Vert \psi\Vert^{2}=\sum_{ik}c_{i}c_{k}N_{ik},
\eeq
with
\beq
N_{ik}=(\pi^{2}/3)(\nu+\beta_{ik})^{-2},
\eeq
where we have defined $\beta_{ik}=\beta_{i}+\beta_{k}$.

The average energy in the state $\psi$ results
\beq
\langle H\rangle=\sum_{ik}c_{i}c_{k}E_{ik}.
\eeq
The matrix $E_{ik}$ is a sum of three terms:
\beq
E_{ik}=\sum_{j=1}^{3}E_{ik}^{(j)},
\eeq
corresponding to the kinetic part ($j=1$), the first term in the interaction
(7) ($j=2$) and to its second term ($j=3$). They are given by
\beq
E_{ik}^{(1)}=(3\pi^{2}/8)\nu^{-1}(\nu+\beta_{ik})^{-1}
(2\nu+3\beta_{ik})^{-1},
\eeq
\beq
E_{ik}^{(2)}=-(3\pi^{2}/4)(\nu+\beta_{ik})^{-1}
(2\nu^{2}+3\nu\beta_{ik}+3\beta_{i}\beta_{k})^{-1},
\eeq
\beq
E_{ik}^{(3)}=(3\pi^{2}/8)(\nu+\beta_{i})^{-1}(\nu+\beta_{k})^{-1}
(2\nu+3\beta_{ik})^{-1}.
\eeq

Numerical calculations show that whatever  the number of gaussians is taken 
in (69)
and however the values of $\beta_{i}$ are chosen, the matrix $E_{ik}$ has no
nonpositive eigenvalue unless all $\beta_{i}=0$, which case corresponds to
the bootstrap solution. Thus the bootstrap solution is the only one with
nonpositive energy in the limit $\nu\rightarrow\infty$. 

 \section{Appendix 2. Monopole part of the Coulomb
interaction}

\ \ \ \ Explicitly the monopole term contribution is given by
\beq
 U_{\alpha_{1},\alpha_{2};\alpha_{3},\alpha_{4}}=16\pi^{2}
\int_{-\infty}^{\infty} dz_{1}\eta_{\alpha_{1}}(z_{1})\psi_{\alpha_{3}}(z_{1})
z_{1}\int_{-\infty}^{z_{1}}\psi_{\alpha_{2}}(z_{2})\eta_{\alpha_{4}}(z_{2})
+(\alpha_{1}\leftrightarrow\alpha_{4},\alpha_{2}\leftrightarrow\alpha_{3}).
\eeq
Here and in the following it is assumed that $l=0$, that is, $l_{1}=l_{3}$
and $l_{2}=l_{4}$. Introduce a function
\beq
\chi_{\alpha_{2},\alpha_{4}}(z)=\int_{-\infty}^{z}dz'
\psi_{\alpha_{2}}(z')\eta_{\alpha_{4}}(z').
\eeq
Once integrating by parts we find
\beq
\chi_{\alpha_{2},\alpha_{4}}(z)=\psi_{\alpha_{2}}(z)\xi_{\alpha_{4}}(z)-
\xi_{\alpha_{2},\alpha_{4}}(z),
\eeq
where the function $\xi_{\alpha_{2},\alpha_{4}}(z)$ with two indices, symmetric
in these, is defined as
\beq
\xi_{\alpha_{2},\alpha_{4}}(z)=\int_{-\infty}^{z}dz'
\xi_{\alpha_{2}}(z')\xi_{\alpha_{4}}(z').
\eeq
As $z\rightarrow\infty$, according to (39),
$\xi_{\alpha_{2},\alpha_{4}}(z)\rightarrow(1/4\pi)
\delta_{\alpha_{2},\alpha_{4}}$, so that
\[\chi_{\alpha_{2},\alpha_{4}}(\infty)=-(1/4\pi)
\delta_{\alpha_{2},\alpha_{4}}.\]
Having this in mind we can rewrite (77) in the form
\beq
 U_{\alpha_{1},\alpha_{2};\alpha_{3},\alpha_{4}}=16\pi^{2}
\int_{-\infty}^{\infty}
dz(\chi_{\alpha_{3},\alpha_{1}}(z)+(1/4\pi)
\delta_{\alpha_{1},\alpha_{3}})'z\chi_{\alpha_{2},\alpha_{4}}(z)
+(\alpha_{1}\leftrightarrow\alpha_{4},\alpha_{2}\leftrightarrow\alpha_{3}).
\eeq
Integrating  by parts, the integral transforms into
\beq
 -16\pi^{2}
\int_{-\infty}^{\infty} dz(\chi_{\alpha_{3},\alpha_{1}}(z)+(1/4\pi)
\delta_{\alpha_{1},\alpha_{3}})(z\chi'_{\alpha_{2},\alpha_{4}}(z)+
\chi_{\alpha_{2},\alpha_{4}}).
\eeq
The term coming from the product $\chi_{\alpha_{3},\alpha_{1}}
z\chi'_{\alpha_{2},\alpha_{4}}$ cancels the contribution
$(\alpha_{1}\leftrightarrow\alpha_{4},\alpha_{2}\leftrightarrow\alpha_{3})$
in (81) so that the monopole contribution becomes
\beq
 U_{\alpha_{1},\alpha_{2};\alpha_{3},\alpha_{4}}=-16\pi^{2}
\int_{-\infty}^{\infty} dz(\chi_{\alpha_{3},\alpha_{1}}(z)
\chi_{\alpha_{2},\alpha_{4}}(z)+(1/4\pi)
\delta_{\alpha_{1},\alpha_{3}}(z\chi'_{\alpha_{2},\alpha_{4}}(z)+
\chi_{\alpha_{2},\alpha_{4}}(z))).
\eeq

Now we substitute the functions $\chi$ in (83) by the symmetric
functions $\xi$ using relation (79). Take the the first term in the integrand
of (83). With (79) we obtain for it 
\[\chi_{\alpha_{3},\alpha_{1}}\chi_{\alpha_{2},\alpha_{4}}=
\psi_{\alpha_{3}}\xi_{\alpha_{1}}\psi_{\alpha_{2}}\xi_{\alpha_{4}}
-\psi_{\alpha_{3}}\xi_{\alpha_{1}}\xi_{\alpha_{2},\alpha_{4}}
-\xi_{\alpha_{3},\alpha_{1}}\psi_{\alpha_{2}}\xi_{\alpha_{4}}
+\xi_{\alpha_{3},\alpha_{1}}\xi_{\alpha_{2},\alpha_{4}}.\]
Having in mind the subsequent symmetrization with respect to the interchange
of gluons 1 and 2, we can change $\alpha_{1}\leftrightarrow
\alpha_{2}$ and $\alpha_{3}\leftrightarrow
\alpha_{4}$ in the second term. Summed with the third term it then gives
\beq
 -\xi_{\alpha_{3},\alpha_{1}}(\psi_{\alpha_{2}}\xi_{\alpha_{4}}+
\xi_{\alpha_{2}}\psi_{\alpha_{4}}).\eeq
Recall now that $\xi_{\alpha_{2}}=(\partial+(1/2)|l_{2}|)\psi_{\alpha_{2}}$
and similarly for $\xi_{\alpha_{4}}$. Integration by parts allows  to
substitute (84) by
\beq
(\xi_{\alpha_{3}}\xi_{\alpha_{1}}-|l_{2}|\xi_{\alpha_{3},\alpha_{1}})
\psi_{\alpha_{2}}\psi_{\alpha_{4}}.
\eeq
So finally the first term in (83) leads to the following three contributions to
the monopole Coulomb energy:
\beq
\tilde{U}^{(1)}_{\alpha_{1},\alpha_{2};\alpha_{3},\alpha_{4}}=-16\pi^{2}
\int_{-\infty}^{\infty} dz\xi_{\alpha_{3},\alpha_{1}}
\xi_{\alpha_{2},\alpha_{4}},
\eeq
\beq
U^{(2)}_{\alpha_{1},\alpha_{2};\alpha_{3},\alpha_{4}}=16\pi^{2}|l_{2}|
\int_{-\infty}^{\infty} dz\xi_{\alpha_{3},\alpha_{1}}
\psi_{\alpha_{2}}\psi_{\alpha_{4}}
\eeq
and
\beq
U^{(3)}_{\alpha_{1},\alpha_{2};\alpha_{3},\alpha_{4}}=-16\pi^{2}
\int_{-\infty}^{\infty} dz(\psi_{\alpha_{3}}\xi_{\alpha_{1}}
\psi_{\alpha_{2}}\xi_{\alpha_{4}}+
\xi_{\alpha_{3}}\xi_{\alpha_{1}}
\psi_{\alpha_{2}}\psi_{\alpha_{4}}).
\eeq
Of these terms the first is  divergent in its present form. It will receive its
meaning after adding  new contributions coming from the rest of the terms
in (83). For that reason we have denoted it with a tilda. 

Now for the rest of the terms in (83).  Changing the function $\chi$ by $\xi$
according to (79) we have under the integral
\[\xi_{\alpha_{2},\alpha_{4}}+z\xi'_{\alpha_{2},\alpha_{4}}=
\psi_{\alpha_{2}}\xi_{\alpha_{4}}-\xi_{\alpha_{2},\alpha_{4}}+
z\psi_{\alpha_{2}}(\partial-(1/2)|l_{2}|)\xi_{\alpha_{4}}.\]
Integration by parts transforms it into
\beq
-\xi_{\alpha_{2},\alpha_{4}}-z\xi_{\alpha_{2}}\xi_{\alpha_{4}}.\eeq
The first term can be combined with (86) to give the final  $U^{(1)}$:
\beq
U^{(1)}_{\alpha_{1},\alpha_{2};\alpha_{3},\alpha_{4}}=16\pi^{2}
\int_{-\infty}^{\infty} dz
\xi_{\alpha_{2},\alpha_{4}}((1/4\pi)\delta_{\alpha_{1},\alpha_{3}}-
\xi_{\alpha_{3},\alpha_{1}}).
\eeq
Now the integral converges due to the property (39). Putting here the explicit
form of the functions $\xi_{\alpha_{i},\alpha_{k}}$ and integrating over $z$
we obtain the term $U^{(1)}$ in its definitive form:
\beq U^{(1)}_{\alpha_{1},\alpha_{2};\alpha_{3},\alpha_{4}}=16\pi^{2}
\int dz_{1}dz_{2}(z_{1}-z_{2})\theta (z_{1}-z_{2})
\xi_{\alpha_{1}}(z_{1})\xi_{\alpha_{3}}(z_{1})
\xi_{\alpha_{2}}(z_{2})\xi_{\alpha_{4}}(z_{4}).
\eeq

 The second term in (89)
gives the last  contribution to the monopole energy
\beq U^{(4)}_{\alpha_{1},\alpha_{2};\alpha_{3},\alpha_{4}}=4\pi
\delta_{\alpha_{1},\alpha_{3}}
\int_{-\infty}^{\infty} dzz
\xi_{\alpha_{2}}\xi_{\alpha_{4}}.
\eeq
This term cancels with the contribution $T^{(2)}$ to the kinetic energy.
Indeed after the Fourier transformation to the $\nu$ space according to (48),
the factor $z$ goes into $i\partial/\partial\nu$. One can then see that (92)
gives exactly the contribution $T^{(2)}$, Eq. (56), with an opposite sign
and with gluons 1 and 2 interchanged, which is of no importance because of the
subsequent symmetrization.

The term $U^{(3)}$ cancels with the part of the 
contact interaction $Q$, Eq. (46), which does not contain factors $\Delta$:
\beq
Q^{(2)}_{\alpha_{1},\alpha_{2};\alpha_{3},\alpha_{4}}=8\pi^{2}
\int_{-\infty}^{\infty} dz(
\psi_{\alpha_{3}}\xi_{\alpha_{1}}\psi_{\alpha_{2}}\xi_{\alpha_{4}}+
\xi_{\alpha_{3}}\xi_{\alpha_{1}}\psi_{\alpha_{2}}\psi_{\alpha_{4}}+
\psi_{\alpha_{3}}\psi_{\alpha_{1}}\xi_{\alpha_{2}}\xi_{\alpha_{4}}+
\xi_{\alpha_{3}}\psi_{\alpha_{1}}\xi_{\alpha_{2}}\psi_{\alpha_{4}}).
\eeq
Summed with $U^{(3)}$ this part gives
\beq (Q^{(2)}+U^{(3)}) _{\alpha_{1},\alpha_{2};\alpha_{3},\alpha_{4}}=8\pi^{2}
\int_{-\infty}^{\infty} dz
(\psi_{\alpha_{1}}\xi_{\alpha_{2}}-\xi_{\alpha_{1}}\psi_{\alpha_{2}})
(\xi_{\alpha_{3}}\psi_{\alpha_{4}}+\psi_{\alpha_{3}}\xi_{\alpha_{4}}).
\eeq
This expression is antisymmetric under the interchange of the gluons 1 and 2
and does not give any contribution to the energy.

So finally the only contributions which remain in the interaction for zero
angular momentum transfer are $U^{(1)}$, $U^{(2)}$ and the part $Q^{(1)}$ of
the contact interaction (46) which contains factors $\Delta$.

\newpage
{\Large\bf References}\\

\noi[1] E. A. Kuraev, L. N. Lipatov and V. S. Fadin, Sov. Phys. JETP {\bf 44}
(1976) 433; {\bf 45} (1977) 199;
Ya. Ya. Balitzky and L. N. Lipatov, Sov. J. Nucl. Phys. {\bf 28} (1978) 822.

\vspace{0.2cm}
\noi[2] E. M. Levin, preprint FERMILAB-CONF-94/068-T (1994).

\vspace{0.2 cm}
\noi[3] M.A.Braun, S.Petersburg  University preprint SPbU-IP-1995/3
(HEP-PH/9502403).

\vspace{0.2cm}
\noi[4] J. Bartels, Nucl. Phys. {\bf B175} (1980) 365; J. Kwiecinski
and M. Praszalowicz, Phys. Lett. {\bf B94} (1980) 413.

\vspace{0.2cm}
\noi[5] L. N. Lipatov, JETP Lett. {\bf 59} (1994) 571.

\vspace{0.2cm}
\noi[6] L. D. Faddeev and G. P. Korchemsky, preprint ITP-SP-14
(HEP-TH/9404173).

\vspace{0.2cm}
\noi[7]  G. P. Korchemsky, preprint HEP-PH/9501232.

\vspace{0.2cm}
\noi[8] P. Gauron, L. N. Lipatov and B. Nicolescu, Phys. Lett. {\bf B260}
(1991) 407.

\vspace{0.2cm}
\noi[9] P. Gauron, L. N. Lipatov and B. Nicolescu, Z.Phys. {\bf C63} (1994)
253.

\vspace{0.2cm}
\noi[10] N. Armesto and M. A. Braun, Santiago Univ. preprint US-FT/9-94
(HEP-TH/9410411).

\vspace{0.2cm}
\noi[11] L. N. Lipatov, Yad. Fiz. {\bf 23} (1976) 642.

\newpage
{\Large\bf Table }
\vspace{0.5cm}\begin{quotation}
Calculated values of the ground state energy per gluon multiplied by 2
(Eq. (31)) for the pomeron ($\epsilon_{2}$) and odderon ($\epsilon_{3}$)  with
different numbers
$r$ of radial functions and angular momenta included. 
\end{quotation}
\vspace{0.2cm}
\begin{center}
\begin {tabular}{|c|c|c|}  \hline
  $r$ & $\epsilon_{2}$ & $\epsilon_{3}$\\\hline
   1  &  0.968        &  0.968     \\\hline
   2  &  0.022        &  0.605     \\\hline
   3  &$-0.475$       &  0.454     \\\hline
   4  &$-0.743$       &  0.379     \\\hline
   5  &$-0.912$       &  0.331     \\\hline
   6  &$-1.032$       &  0.298     \\\hline
      &               &            \\\hline
$\infty$& $-2.773$    &            \\\hline
\end{tabular}
\end{center}
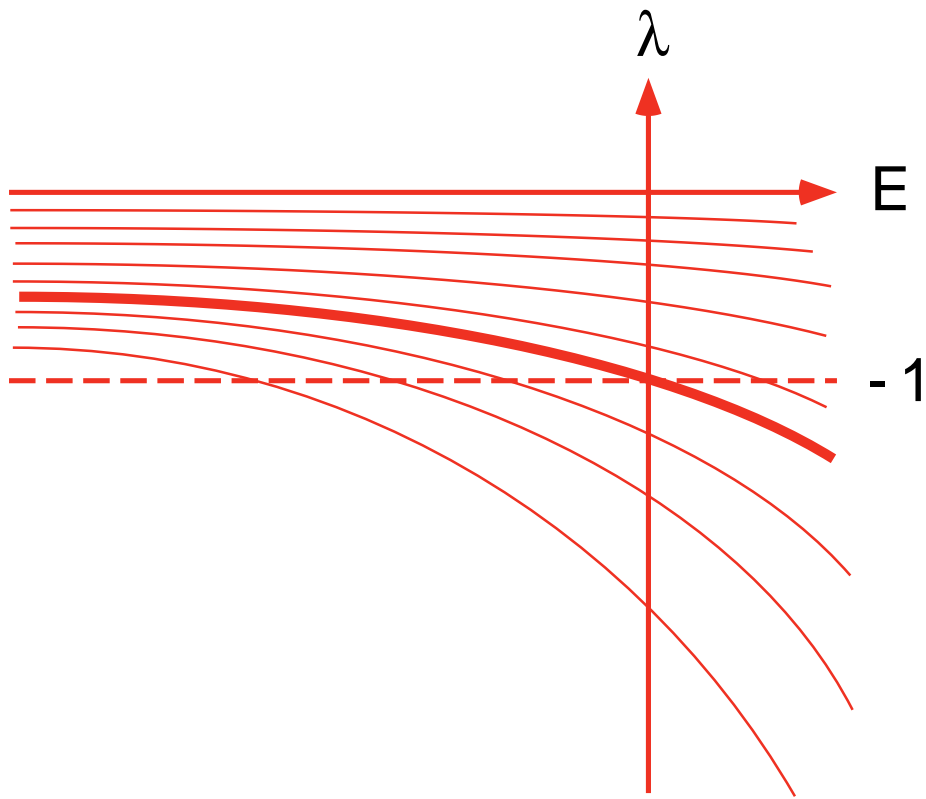
\end{document}